\documentclass[abstract,longauth]{aa}
\usepackage{graphicx}
\usepackage{txfonts}
\usepackage{natbib}

\def\HI{H{\,\small I}}

\newcommand{\mJybeam}{mJy beam$^{-1}$}
\newcommand{\mJybeamchan}{mJy beam$^{-1}$chan$^{-1}$}

\newcommand{\kms}{$\,$km$\,$s$^{-1}$}
\newcommand{\whz}{WHz$^{-1}$}

\bibpunct{(}{)}{;}{a}{}{,}

\begin{document}

\title{The peculiar radio galaxy 4C~35.06: a case for recurrent AGN activity?}

\titlerunning{Recurring AGN activity in 4C~35.06}
\authorrunning{Shulevski et al.}	
\author{A.~Shulevski\inst{1,2}\and
	R.~Morganti\inst{1,2}\and
	P.~D.~Barthel\inst{1}\and
	M.~Murgia\inst{3}\and
	R.~J.~van~Weeren\inst{4}\and
	G.~J.~White\inst{5,6}\and
	M.~Br{\"u}ggen\inst{7}\and
	M.~Kunert-Bajraszewska\inst{8}\and
	M.~Jamrozy\inst{9}\and
	P.~N.~Best\inst{10}\and
	H.~J.~A.~R\"{o}ttgering\inst{11}\and
	K.~T.~Chyzy\inst{9}\and
	F.~de~Gasperin\inst{7}\and
	L.~B\^{\i}rzan\inst{7}\and
	G.~Brunetti\inst{40}\and
	M.~Brienza\inst{2,1}\and
	D.~A.~Rafferty\inst{7}\and
	J.~Anderson\inst{12}\and 
	R.~Beck\inst{13}\and 
	A.~Deller\inst{2}\and 
	P.~Zarka\inst{39}\and
	D.~Schwarz\inst{36}\and 
	E.~Mahony\inst{2}\and
	E.~Orr\'{u}\inst{2}\and
	M.~E.~Bell\inst{14}\and
	M.~J.~Bentum\inst{2,15}\and 
	G.~Bernardi\inst{4}\and 
	A.~Bonafede\inst{7}\and 
	F.~Breitling\inst{16}\and
	J.~W.~Broderick\inst{17,18}\and
	H.~R.~Butcher\inst{19}\and 
	D.~Carbone\inst{20}\and 
	B.~Ciardi\inst{21}\and 
	E.~de Geus\inst{2,22}\and
	S.~Duscha\inst{2}\and 
	J.~Eisl\"offel\inst{23}\and
	D.~Engels\inst{7}\and 
	H.~Falcke\inst{24,2}\and
	R.~A.~Fallows\inst{2}\and
	R.~Fender\inst{17}\and
	C.~Ferrari\inst{41}\and
	W.~Frieswijk\inst{2}\and 
	M.~A.~Garrett\inst{2,11}\and 
	J.~Grie\ss{}meier\inst{25,26}\and
	A.~W.~Gunst\inst{2}\and 
	G.~Heald\inst{2,1}\and 
	M.~Hoeft\inst{23}\and 
	J.~H\"orandel\inst{24}\and
	A.~Horneffer\inst{14}\and 
	A.J.~van~der~Horst\inst{20}\and
	H.~Intema\inst{11,27}\and 
	E.~Juette\inst{28}\and
	A. ~Karastergiou\inst{17}\and
	V.~I.~Kondratiev\inst{2,29}\and
	M.~Kramer\inst{13,30}\and 
	M.~Kuniyoshi\inst{31}\and 
	G.~Kuper\inst{2}\and 
	P.~Maat\inst{2}\and 
	G.~Mann\inst{17}\and 
	R.~McFadden\inst{2}\and
	D.~McKay-Bukowski\inst{32,33}\and
	J.~P.~McKean\inst{2,1}\and 
	H.~Meulman\inst{2}\and 
	D.~D.~Mulcahy\inst{18}\and 
	H.~Munk\inst{2}\and 
	M.~J.~Norden\inst{2}\and
	H.~Paas\inst{34}\and 
	M.~Pandey-Pommier\inst{35}\and 
	R.~Pizzo\inst{2}\and 
	A.~G.~Polatidis\inst{2}\and 
	W.~Reich\inst{13}\and
	A.~Rowlinson\inst{14}\and
	A.~M.~M.~Scaife\inst{18}\and
	M.~Serylak\inst{17}\and 
	J.~Sluman\inst{2}\and
	O.~Smirnov\inst{37,38}\and
	M.~Steinmetz\inst{16}\and 
	J.~Swinbank\inst{20}\and 
	M.~Tagger\inst{25}\and 
	Y.~Tang\inst{2}\and 
	C.~Tasse\inst{39}\and
	S.~Thoudam\inst{24}\and 
	M.~C.~Toribio\inst{2}\and 
	R.~Vermeulen\inst{2}\and
	C.~Vocks\inst{16}\and 
	R.~A.~M.~J.~Wijers\inst{20}\and 
	M.~W.~Wise\inst{2,20}\and 
	O.~Wucknitz\inst{13}
	}
	
	\institute{University of Groningen, Kapteyn Astronomical Institute, Landleven 12, 9747 AD Groningen, The Netherlands\\
		   \email{pdb@astro.rug.nl}\and
		   ASTRON, the Netherlands Institute for Radio Astronomy, Postbus 2, 7990 AA, Dwingeloo, The Netherlands\\
		   \email{shulevski@astron.nl, morganti@astron.nl}\and
		   INAF - Osservatorio Astronomico di Cagliari, Via della Scienza 5, I-09047, Selargius (Cagliari)\and
		   Harvard-Smithsonian Center for Astrophysics, 60 Garden Street, Cambridge, MA 02138, USA\and
		   Department of Physics and Astronomy, The Open University, Milton Keynes MK7 6AA, England\and
		   RAL Space, The Rutherford Appleton Laboratory, Chilton, Didcot, Oxfordshire OX11 0QX, England\and
		   Universit\"{a}t Hamburg, Hamburger Sternwarte, Gojenbergsweg 112, D-21029, Hamburg, Germany\and
		   Toru\'n Centre for Astronomy, Faculty of Physics, Astronomy and Informatics, NCU, Grudziacka 5, 87-100 Toru\'n, Poland\and
		   Obserwatorium Astronomiczne, Uniwersytet Jagiello\'{n}ski, ul Orla 171, 30-244, Krak\'{o}w, Poland\and
		   SUPA, Institute for Astronomy, Royal Observatory Edinburgh, Blackford Hill, Edinburgh EH9 3HJ, UK\and
		   Leiden Observatory, Leiden University, PO Box 9513, 2333 RA Leiden, The Netherlands\and
		   Helmholtz-Zentrum Potsdam, DeutschesGeoForschungsZentrum GFZ, Department 1: Geodesy and Remote Sensing, Telegrafenberg, A17, 14473 Potsdam, Germany\and
		   Max-Planck-Institut f\"{u}r Radioastronomie, Auf dem H\"ugel 69, 53121 Bonn, Germany\and
		   CSIRO Australia Telescope National Facility, PO Box 76, Epping NSW 1710, Australia\and
		   University of Twente, The Netherlands\and
		   Leibniz-Institut f\"{u}r Astrophysik Potsdam (AIP), An der Sternwarte 16, 14482 Potsdam, Germany\and
		   Astrophysics, University of Oxford, Denys Wilkinson Building, Keble Road, Oxford OX1 3RH\and
		   School of Physics and Astronomy, University of Southampton, Southampton, SO17 1BJ, UK\and
		   Research School of Astronomy and Astrophysics, Australian National University, Mt Stromlo Obs., via Cotter Road, Weston, A.C.T. 2611, Australia\and
		   Anton Pannekoek Institute, University of Amsterdam, Postbus 94249, 1090 GE Amsterdam, The Netherlands\and
		   Max Planck Institute for Astrophysics, Karl Schwarzschild Str. 1, 85741 Garching, Germany\and
		   SmarterVision BV, Oostersingel 5, 9401 JX Assen\and
		   Th\"{u}ringer Landessternwarte, Sternwarte 5, D-07778 Tautenburg, Germany\and
		   Department of Astrophysics/IMAPP, Radboud University Nijmegen, P.O. Box 9010, 6500 GL Nijmegen, The Netherlands\and
		   LPC2E - Universite d'Orleans/CNRS\and
		   Station de Radioastronomie de Nancay, Observatoire de Paris - CNRS/INSU, USR 704 - Univ. Orleans, OSUC , route de Souesmes, 18330 Nancay, France\and
		   National Radio Astronomy Observatory, 520 Edgemont Road, Charlottesville, VA 22903-2475, USA\and
		   Astronomisches Institut der Ruhr-Universit\"{a}t Bochum, Universitaetsstrasse 150, 44780 Bochum, Germany\and
		   Astro Space Center of the Lebedev Physical Institute, Profsoyuznaya str. 84/32, Moscow 117997, Russia\and
		   Jodrell Bank Center for Astrophysics, School of Physics and Astronomy, The University of Manchester, Manchester M13 9PL,UK\and
		   National Astronomical Observatory of Japan, Japan\and
		   Sodankyl\"{a} Geophysical Observatory, University of Oulu, T\"{a}htel\"{a}ntie 62, 99600 Sodankyl\"{a}, Finland\and
		   STFC Rutherford Appleton Laboratory,  Harwell Science and Innovation Campus,  Didcot  OX11 0QX, UK\and
		   University of Groningen, Center for Information Technology (CIT), The Netherlands\and
		   Centre de Recherche Astrophysique de Lyon, Observatoire de Lyon, 9 av Charles Andr\'{e}, 69561 Saint Genis Laval Cedex, France\and
		   Fakult\"{a}t f\"{u}r Physik, Universit\"{a}t Bielefeld, Postfach 100131, D-33501, Bielefeld, Germany\and
		   Department of Physics and Elelctronics, Rhodes University, PO Box 94, Grahamstown 6140, South Africa\and
		   SKA South Africa, 3rd Floor, The Park, Park Road, Pinelands, 7405, South Africa\and
		   LESIA, UMR CNRS 8109, Observatoire de Paris, 92195 Meudon, France\and
		   IRA-INAF, via P. Gobetti 101, 40129 Bologna, Italy\and
		   Laboratoire Lagrange, UMR7293, Universit\`{e} de Nice Sophia-Antipolis, CNRS, Observatoire de la C\'{o}te d'Azur, 06300 Nice, France}
		   
\date{\today}
	
\abstract
	{Using observations obtained with the LOw Fequency ARray (LOFAR), the Westerbork Synthesis Radio Telescope (WSRT) and archival Very Large Array (VLA) data, we have traced the radio emission to large scales in the complex source 4C~35.06 located in the core of the galaxy cluster Abell~407. At higher spatial resolution ($ \sim 4\arcsec $), the source was known to have two inner radio lobes spanning 31~kpc and a diffuse, low-brightness extension running parallel to them, offset by about 11~kpc (in projection).
	
	At 62~MHz, we detect the radio emission of this structure extending out to 210~kpc. At 1.4~GHz and intermediate spatial resolution ($ \sim 30\arcsec $), the structure appears to have a helical morphology.

	We have derived the characteristics of the radio spectral index across the source. We show that the source morphology is most likely the result of at least two episodes of AGN activity separated by a dormant period of around 35 Myr. The outermost regions of radio emission have a steep spectral index ($ \alpha < -1 $), indicative of old plasma. We connect the spectral index properties of the resolved source structure with the integrated flux density spectral index of 4C~35.06 and suggest an explanation for its unusual integrated flux density spectral shape (a moderately steep power law with no discernible spectral break), possibly providing a proxy for future studies of more distant radio sources through inferring their detailed spectral index properties and activity history from their integrated spectral indices.

	The AGN is hosted by one of the galaxies located in the cluster core of Abell~407. We propose that it is intermittently active as it moves in the dense environment in the cluster core. In this scenario, the AGN turned on sometime in the past, and has produced the helical pattern of emission, possibly a sign of jet precession/merger during that episode of activity. Using LOFAR, we can trace the relic plasma from that episode of activity out to greater distances from the core than ever before.

	Using the the WSRT, we detect \HI\ in absorption against the center of the radio source. The absorption profile is relatively broad (FWHM of 288 \kms), similar to what is found in other clusters. The derived column density is $ N_{\mathrm{HI}} \, \sim 4 \, \cdot \, 10^{20} $ cm$ ^{-2} $ for a $ T_{\mathrm{spin}} \, = \, 100 $ K. This detection supports the connection - already suggested for other restarted radio sources - between the presence of cold gas and restarting activity. The cold gas appears to be dominated by a blue-shifted component although the broad \HI\ profile could also include gas with different kinematics.

	Understanding the duty cycle of the radio emission as well as the triggering mechanism for starting (or restarting) the radio-loud activity can provide important constraints to quantify the impact of AGN feedback on galaxy evolution. The study of these mechanisms at low frequencies using morphological and spectral information promises to bring new important insights in this field.}

\maketitle

\section{Introduction}
\label{c3:intro}

For some time now it has been recognized that the active phase of radio-loud active galactic nuclei (AGN) can be recurrent \citep[see][for a review]{RefWorks:6}. Therefore, this activity can have a repeated impact on the AGN host galaxy and its environment, starting from the initial phase of the radio source's life \citep{RefWorks:8} on to its later stages \citep{RefWorks:153}.
The morphology of a (recurrent) radio source can vary significantly \citep[see e.g.][]{RefWorks:28, RefWorks:34}. Therefore, only a combination of radio morphology and radio spectral index\footnote{We use the $ S \propto \nu ^{\alpha} $ definition for the spectral index $ \alpha $ throughout this writing.} analyses can reliably tell us whether a radio source had recurring activity, thereby affecting its host galaxy more than once. A typical radio-loud AGN in its active phase (which can last up to several $ 10^{7} $ years) is characterized by the presence of a radio core and extended lobes with or without visible jets. When the AGN shuts down, emission from the core disappears for the duration of the "off" phase. The large scale radio lobes might still be replenished with particles if the information of the core shutdown has not had enough time to propagate out. However, once the supply of energetic particles coming from the core stops, the lobe plasma starts to age producing a so-called {\sl relic} structure and the radio source enters a fading phase in its life-cycle. It is possible that during this time the AGN re-activates, in which case we can again observe an active radio core (possibly with new jets / lobes expanding outward into the ISM / IGM) co-existing with a fading radio relic on larger scales.

The energy loss mechanism for relic radio plasma is mostly through synchrotron radiation and inverse Compton (IC) scattering off of cosmic microwave (CMB) photons (with possible adiabatic expansion related losses). Its observational signature is a steepening of the radio spectrum of the emission regions (synchrotron radiation and IC losses deplete high-energy particles faster than low-energy ones), making them relatively brighter at low frequencies \citep{RefWorks:126}. Thus, the relic radio emission is typically faint and detecting it requires low-frequency observations with high sensitivity.

This work is part of a larger study where we investigate radio sources with restarted activity expanding our studies to low frequencies, using the LOw Frequency ARray \citep[LOFAR, ][]{RefWorks:157}. Our aim is to describe in more detail their activity history (duty cycle). We also benefit from the advantage of having sufficient spatial resolution to be able to study resolved source structure at low frequencies. We relate our results to the integrated flux density spectral properties of these sources in an attempt to extrapolate our results to the high redshift source population (in which case we would have poorer spatial resolution).

\begin{figure}
  \centering
  \includegraphics[width=0.5\textwidth]{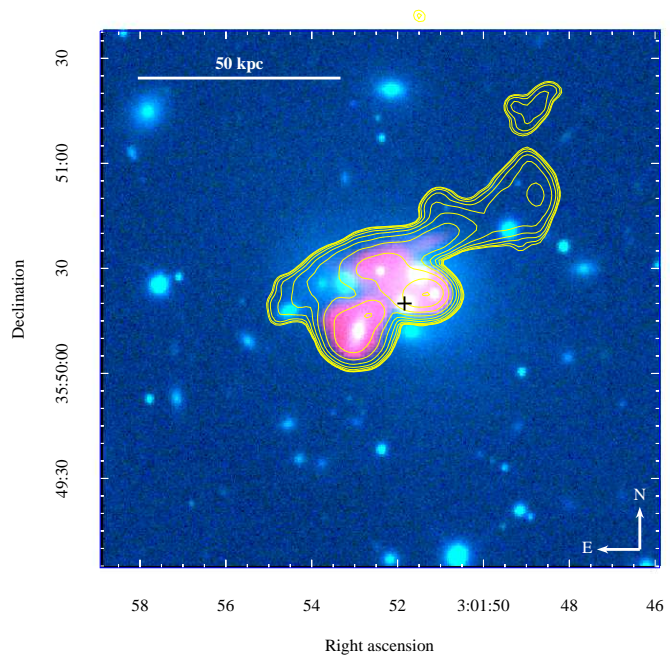}
  \caption{Optical image of UGC~2489 taken from the Sloan Digital Sky Survey \citep[SDSS,][]{RefWorks:240} showing its multi-component core structure. The red intensity overlay shows a 4885 MHz VLA archival image. In yellow we over-plot contours from a 1415 MHz VLA archival image. Contours are log scaled, there are 10 contour levels going from $ 4\sigma $ to $ 500 \sigma $.  Details about the VLA images can be found in Table \ref{c3:table:1}. The stellar halo around the cores is roughly 40 kpc in extent. The position of the core hosting the AGN is shown with a black cross.}
\label{c3:VLA_in}
\end{figure}

In this work, we investigate the complex radio morphology of 4C~35.06 \citep[B2~0258+35B;][]{RefWorks:112}. \cite{RefWorks:110} did a study of UGC~2489 using the Very Long Baseline Array (VLBA) and they detected an extended source on a milli arcsec scale with a core flux density of $ \sim 2.6 $ mJy, stating that the VLBA detection coincides with one of the cores of the cD galaxy UGC~2489 (Figure \ref{c3:VLA_in}). UGC~2489 represents the core of the Abell 407 galaxy cluster \citep[][see Figure \ref{c3:VLA_in}]{RefWorks:111}. The radio luminosity of 4C~35.06  is $ L_{\mathrm{1.4GHz}} \, = \, 2.5 \, \cdot \, 10^{24} $ \whz. This value is consistent according to \cite{RefWorks:150} with being a low excitation radio galaxy (LERG) as confirmed by the low-resolution optical spectra of the host galaxy \citep{RefWorks:111, RefWorks:154} showing no emission lines.

In the VLA maps we can see two inner radio lobes spanning $ 31 $ kpc and a diffuse, low-brightness extension running parallel to them, oriented southeast to northwest (see Figure \ref{c3:VLA_in}). The puzzling structure hints at the possibility of multiple AGN activity episodes combined with motion of the host core within the cD galaxy, making this source particularly interesting for studying AGN recurrent activity and the triggering of the active phases.

We made use of images available in the Jansky Very Large Array (VLA) archive. Furthermore, we took advantage of the fact that this source was included in one of the fields observed by the LOFAR during commissioning tests.

In addition, we also present new Westerbork Synthesis Radio Telescope (WSRT) observations obtained to investigate the presence of cold gas (\HI) in this object. As mentioned above, one of the open questions is the relation between the gas and the fueling of multiple phases of activity. Atomic neutral hydrogen can, in principle, provide such fuel.

The paper is organized in the following way. Section \ref{c3:obsdata} discusses the data reduction procedure, while in Section \ref{c3:res} we present the LOFAR image and quantities derived from it. We discuss the implications of our results, in combination with results of previous studies in Section \ref{c3:disc}, and we conclude with a summary in Section \ref{c3:fin}.

Throughout this paper we assume a flat $ \Lambda $ cold dark matter ($ \Lambda CDM $) cosmology with H$ _{\mathrm{0}} $ = 73 km s$ ^{-1} $ Mpc$ ^{-1} $, $ \Omega_{\mathrm{\Lambda}} = 0.73 $ and $ \Omega_{\mathrm{m}} = 0.27 $. At the distance of 4C~35.06, using $ z =  0.046276 $ (corrected to the reference frame defined by the CMB), this results in 1 \arcsec\ = 0.872 kpc \citep{RefWorks:155}.

\begin{table}[htpb]
\noindent \caption{\small Summary of the images of 4C~35.06 used in this work.}
\label{c3:table:1}
\small
\begin{tabular}{p{1.5cm} p{1.5cm} p{1.5cm} p{2cm} p{0.5cm}} 
\hline\hline
Instrument & $ \nu $ [MHz] & Beam size & Noise [mJy/b] & Ref.\\
\hline
LOFAR & 61 & 51\arcsec x 43\arcsec & 35 & 1\\
WSRT & 1360 & 30\arcsec x 13\arcsec & 0.18 & 1\\
VLA (B) & 1415 & 4\farcs8 x 4\farcs1 & 0.29 & 2\\
VLA (C) & 4885 & 3\farcs8 & 0.04 & 3\\
\hline 
\end{tabular}
\tablebib{(1) This work; (2) \citep[][VLA archive image]{RefWorks:120}; (3) \cite[][VLA archive image]{RefWorks:112}.}
\end{table}

\section{Data available, new observations and data reduction}
\label{c3:obsdata}

\subsection{LOFAR}
\label{c3:lofar}

The observations using LOFAR were carried out in the night of October 8, 2011 using the low band antennas (LBA) at 62 MHz centered on B2~0258+35A observed as part of the LOFAR commissioning efforts. 4C~35.06 is located about half a degree to the north from the phase center. Table \ref{c3:table:2} summarizes the observational configuration.

\begin{table}[!htpb]
\noindent \caption{\small LOFAR observational configuration}
\label{c3:table:2}
\small
\begin{tabular}{ p{4cm} p{4cm}}
\hline\hline
No. of sub-bands & 28\\
Central Frequency [MHz] & 62 \\
Bandwidth [MHz] & 5.6 \\
Integration time & 1 second\\
Observation duration & 6 hours\\
Polarization & Full Stokes \\
No. of LOFAR stations used & 32\\
$ UV $ range & $ 20 \lambda - 4000 \lambda $\\
\hline
\end{tabular}
\end{table}

The data were flagged to remove radio frequency interference (RFI) using the AO flagger package \citep{RefWorks:133} and averaged to a frequency resolution of 192 kHz per channel and a time resolution of 10 seconds per sample. Calibration was performed using the Black Board Selfcal (BBS) package using the complete bandwidth. The set up of the observations did not include observations of a flux calibrator, therefore the initial calibration model was derived from a VLSS\footnote{VLSS \citep[][]{RefWorks:128} is the VLA Low frequency Sky Survey carried out at a frequency of 74 MHz} image following the procedure described in \cite{RefWorks:156} and \cite{RefWorks:158}. The image from which the initial model was derived had a size of $ 8^{\circ}\times8^{\circ} $ to encompass the full width at half maximum (FWHM) of the LOFAR station (primary) beam, and to ensure that any brighter sources beyond the FWHM are included in the model. The model components were specified as having a spectral index of $ \alpha \, = \, -0.8  $. During the calibration, we have solved for the complex valued station gains and for the differential total electron content (dTEC) over the full bandwidth.

Imaging was done using the Common Astronomy Software Applications package \citep[CASA;][]{RefWorks:175}, using multi-scale clean \citep{RefWorks:174}. We imaged by selecting the data in $ UV $ distance below 20 km, limiting ourselves to stations having satisfactory calibration solutions, avoiding as much as possible calibration errors induced by the ionosphere, while retaining enough spatial resolution. We weighted the baselines used for imaging using Briggs weights with the robust parameter set to $ -0.1 $ \citep{RefWorks:185}. We have used multi-frequency synthesis (assuming no frequency dependence of the sky model) imaging over the entire bandwidth.

After obtaining the initial image, we have performed a correction for the primary (station) beam shape to ensure a correct flux scale over the FoV. The correction was applied by deriving the average LOFAR station beam, using CASA to image (with the same settings as were applied in the imaging) a grid of BBS simulated point sources covering the LOFAR FoV. The recovered flux densities were used to derive the primary beam which we use to correct the image.

Before the final imaging, three self-calibration runs were performed by iterating over the procedure described above. We have reached an image noise of about 35 \mJybeam\ measured away from bright sources. The size of the restoring beam is  51\arcsec x 43\arcsec.\\
We have checked the flux scale across the final image, by comparing the flux densities of sources in our LOFAR image with their values in existing surveys. Compact sources with $ S/N > 10 $ were extracted using the PyBDSM package \citep{RefWorks:241} and their flux density at 62~MHz compared with the  extrapolation from the VLSS, WENSS\footnote{WENSS stands for the Westerbork Northern Sky Survey carried out at a frequency of 325 MHz \citep{RefWorks:138}} and NVSS\footnote{NVSS stands for the NRAO VLA Sky Survey carried out at a frequency of 1.4 GHz \citep{RefWorks:139}} catalogs. The derived LOFAR flux densities were estimated to be accurate to $ \sim 10 $\% level.

A check of the positional accuracy of the LOFAR image was performed by plotting the differences between positions of the sources extracted from the LOFAR image and matched sources from the NVSS survey. Again, we used PyBDSM to extract point sources from the LOFAR image. Positional differences for source matches between the WENSS and NVSS catalogs were used as control. On average, the derived position offset for point sources in the vicinity of 4C~35.06 extracted from the LOFAR image and the corresponding NVSS catalog sources is less than 5\arcsec\  without any obvious systematics. Thus, for the purpose of this study, we concluded that there is no position offset in the LOFAR image.

\subsection{\HI\ - 21cm and continuum from the WSRT}
\label{c3:hi-obs}

WSRT observations to search for \HI\ in 4C~35.06 were performed at three epochs (2013 April 8 and 11, June 12) for  8, 7 and 6 hours respectively. The observing band was centered on 1356.8 MHz, the frequency of  \HI\ for z = 0.0464,  the redshift of 4C~35.06. The total bandwidth was 20 MHz, corresponding to a velocity range of $ \sim 4422 $ \kms\, and was covered with 1024 channels (dual polarization). In order to increase the signal-to-noise ratio, four channels were binned resulting in a spectral resolution of 17 \kms\ in the output data cube. The data reduction was performed using the MIRIAD package \citep{RefWorks:63}. After flagging, bandpass and phase calibration were performed. The continuum emission was subtracted, using the task \textit{uvlin}, by making a second-order polynomial fit to the line-free channels of each visibility record.

Cubes with different weighting were obtained. For an intermediate weighting (Briggs with the robust parameter set to 0.4), the final r.m.s. noise was 0.2 \mJybeamchan\ for a velocity resolution of 15 \kms\ and spatial resolution of 34\arcsec\ x 14\arcsec.
In addition to the line cube, a continuum image was obtained using the line-free channels. The peak of the continuum emission was 221 mJy and the r.m.s. noise was 0.18 \mJybeam.

\begin{figure*}[htpb]
\centering
\centering \includegraphics[width=180mm, angle=0]{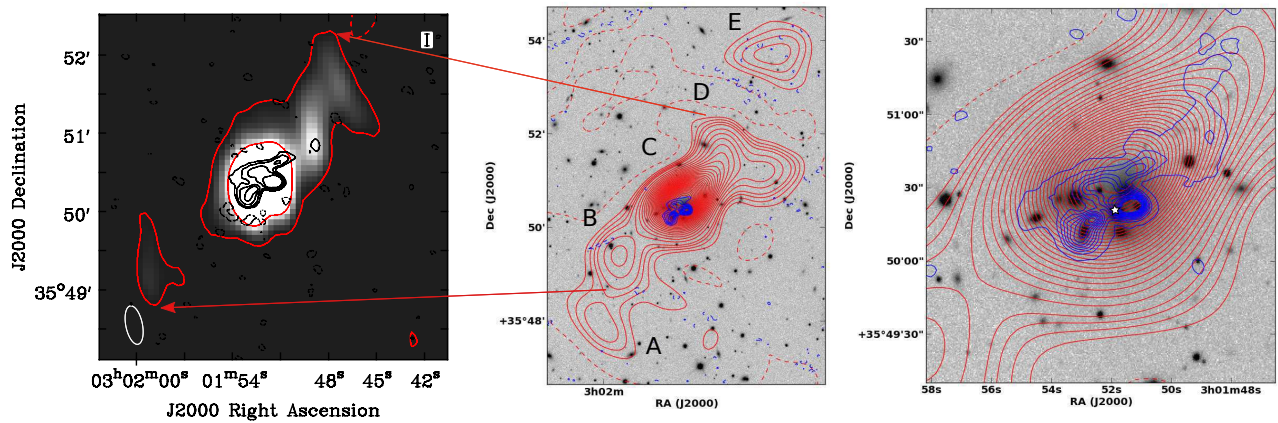}
\caption{\textbf{Left:} WSRT 1360 MHz continuum image of 4C~35.06. Overlaid in red are contours taken from the same image, with levels $ (-5, 15, 150)\sigma $, while plotted in black are 4885 MHz VLA contours of the central region with contour levels of $ (-5, 50, 100, 200, 350)\sigma $. The WSRT beam is shown in the lower left corner. \textbf{Middle:} The 4C 35.06 radio source as seen by LOFAR at 62 MHz, represented by red contours (33 levels spanning $ (-3, 183)\sigma $) over an inverted grayscale SDSS image of the central region of the Abell 407 cluster. The peak flux density in this map is 9.1 Jy. Regions in the LOFAR map are labeled A through E. Overlaid in blue are the VLA 4885 MHz image contours. \textbf{Right:} LOFAR 62 MHz (red) and VLA 1415 MHz (blue) contours overlaid over an inverted grayscale SDSS image of the cluster core (UGC~2489). The location of the core hosting the AGN is indicated with a white star. Arrows show the extent and the location of the regions of interest in different images. The relevant beam sizes and noise levels of the images are given in Table \ref{c3:table:1}.}
\label{c3:panel}
\end{figure*}

\section{Results}
\label{c3:res}

\subsection{Radio continuum structure from WSRT and LOFAR}
\label{c3:reslofar}

Our LOFAR and WSRT images of 4C~35.06, as well as data derived from the archival VLA images are presented in Fig. \ref{c3:panel}.
Looking at the morphology traced by the different images at different resolution, we can note several things.

Firstly, Figure \ref{c3:panel} indicates that the emission parallel to the inner radio lobes is connected to the more distant regions through a helical structure (best seen in the left panel of Figure \ref{c3:panel} as a bent tail of emission extending to the northwest from the central region, and then turning to the south; we note that there is also a similar almost symmetric extension to the southeast). The WSRT image has sufficiently high resolution to enable us to justify the claim that the helicity is real and not an artifact of the elongated beam. It has recently been imaged by \cite{biju} using the Giant Metre Wave Radio Telescope (GMRT). Observations using LOFAR at a frequency of 150 MHz and higher resolution than described in this work also show the helical substructure (Marisa Brienza, private communication). The helicity might indicate precession in the radio jet which has deposited the plasma, triggered by accretion disk instabilities \citep{RefWorks:176} or maybe a possible sign of a past black hole merger of two cores in UGC~2489.

The zoom-in of the central region of 4C~35.06 in the LOFAR image (labeled C) is shown in the right panel of Figure \ref{c3:panel} along with the same region as observed by the VLA at higher frequencies.

The source region labeled E is noticeable as a barely visible surface brightness enhancement in a VLSS image of 4C~35.06, and it is detected at a 2$ \sigma $ level in a VLA image by \cite{RefWorks:120}. It is now unambiguously detected in the LOFAR image. Its projected distance from the central region (C) is $ \sim $ 210 kpc.

The region labeled A was previously detected by the VLSS (blended with the rest of the source) and partially by the WENSS, however our LOFAR map detects it with better resolution and allows us to clearly morphologically separate it from other regions of 4C~35.06.

The second interesting result is that, as shown in the right panel of Figure \ref{c3:panel}, the peak flux of 4C~35.06 as seen by LOFAR is offset by 10\arcsec\ to the northwest with respect to the radio lobes seen in the VLA maps. This is a sufficiently large value to be considered a real effect and not a systematic offset (see Sect. \ref{c3:lofar}). We interpret this difference between the peak flux density position and overall source position as intrinsic in the case of the extended source 4C~35.06 and indicative of its (spectral) properties.

\subsection{\HI\ gas}
\label{c3:resHI}

\begin{figure}
  \centering
  \includegraphics[width=95mm]{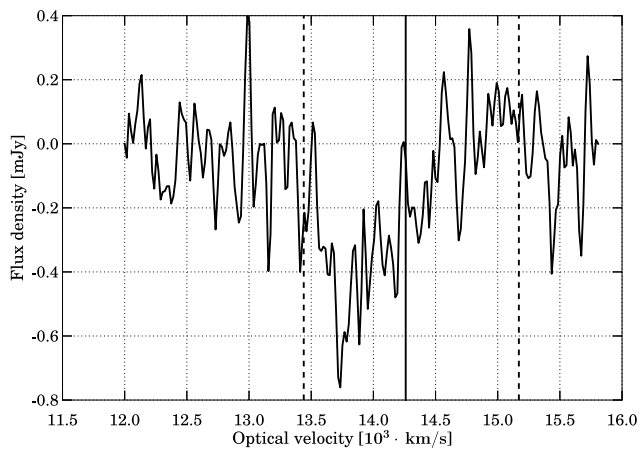}
  \caption{\HI\ absorption profile in 4C~35.06 (Hanning smoothing applied). The full vertical line indicates the systemic velocity of 4C~35.06 host galaxy, while the two vertical dashed lines show the highest and lowest velocity of the galaxies in the cluster core.}
  \label{c3:hi}
\end{figure}

Atomic neutral hydrogen is detected against the central region of 4C~35.06. Figure \ref{c3:hi} shows the \HI\ profile (obtained from the combination of all WSRT data) extracted against the brighter (western) lobe of the central source (Figure \ref{c3:VLA_in}). The profile is obtained after Hanning smoothing and averaged over a region encompassing the western inner lobe. Given the limited spatial resolution of the WSRT \HI\ observations, we can only say that \HI\ is seen in from of the western lobe which, in fact, includes the radio core of the source.

The profile is complex, with a broader component with FWHM of 288 \kms\ centered at 13858 \kms. The peak absorption is $ \sim 0.8 $ mJy which corresponds to a low optical depth of $ \tau \, = \, 0.0036 $. From the parameters of the absorption we can estimate a column density of $ N_{\mathrm{HI}} \, \sim 4 \, \cdot \, 10^{20} $ cm$ ^{-2} $ for a $ T_{\mathrm{spin}} \, = \, 100 $ K.

The \HI\ profile is broader than that typically found in radio galaxies \citep{RefWorks:249}, but resembles what is observed in some other clusters; we discuss this in more details in Section \ref{c3:coldgas}. While at uniform weighting the profile is dominated by two relatively narrow components, the broad component becomes more prominent at lower spatial resolution, suggesting that it belongs to a diffuse distribution of gas. However, given the low S/N of the detection this result will need to be verified with deeper observations.
Compared to the systemic velocity of the host galaxy (14120 \kms\ ), the peak of the \HI\ profile appears blueshifted (see Figure \ref{c3:hi}). However, the total width of the profile encompasses the range in velocity of the galaxies in the cluster core.

In addition to the absorption profile, there are at least three detections of \HI\ emission in the WSRT FoV. All three have an optical counterpart in the SDSS. One of these detections is identified with UGC~2493, a (disturbed, possibly merging) galaxy pair having a redshift of z=0.0437, which makes it a member of Abell~407 according to \cite{RefWorks:119}. It is located at a radial distance of 210 kpc from the Abell~407 cluster core (UGC~2489). The other two detections are located at coordinates: J030219+354601 (\HI\ disk associated with a edge on galaxy) and J030217+355533.

\subsection{Spectral analysis}
\label{c3:specan}

Following the work of \cite{RefWorks:34}, we investigate the characteristics of the integrated spectrum of the source and then we turn to the spatially resolved analysis of the spectral index with the aim of getting a detailed handle on the past and present activity history.

\begin{figure}[htpb]
  \centering
  \includegraphics[width=0.5\textwidth]{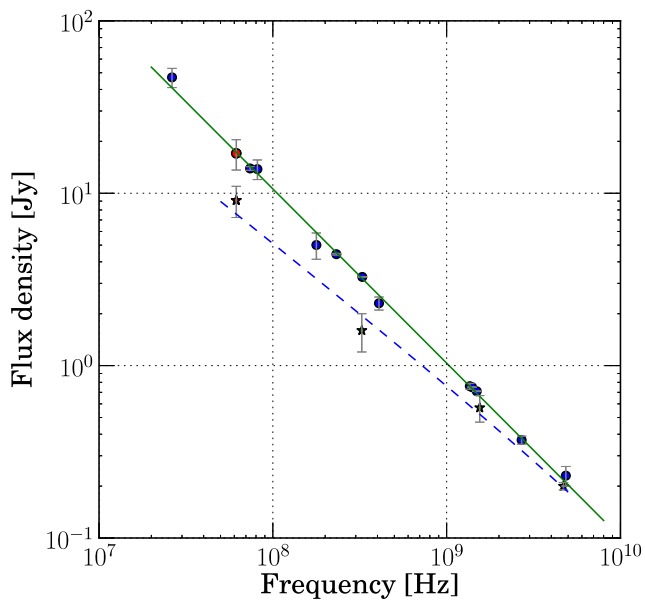}
  \caption{Integrated flux density values for 4C~35.06 based on the data listed in Table \ref{c3:table:global} (circles) and its central region (stars). The LOFAR data points are in red. The green line represents a power law fit, while the blue dashed line gives the synchrotron aging model fit for region C.}
\label{c3:spix}
\end{figure}

We have compiled integrated flux density measurements of 4C~35.06, as shown in Table \ref{c3:table:global}. The flux density values are taken from the literature, or measured from archival images, with comparable sensitivity to the extended emission. They are plotted in Figure \ref{c3:spix} along with our LOFAR flux value.

\begin{table}[htpb!]
\noindent \caption{\small Data used in Figure \ref{c3:spix}.}
\label{c3:table:global}
\small
\begin{tabular}{p{2cm} p{1.2cm} p{2cm} p{2cm}}     
\hline\hline\\
Source & $ \nu $ [MHz] & $ S_{\mathrm{int}} $ [Jy] & $ S_{\mathrm{C}} $ [Jy] \\
\hline\\
Clark Lake\tablefootmark{a} & 26.3 & $ 47.0 \pm 6.0 $ & - \\
LOFAR\tablefootmark{b} &   61.6 & $ 17.04 \pm 3.41 $ & $ 9.1 \pm 1.9 $ \\
VLSS\tablefootmark{c}  &   74.0 & $ 13.92 \pm 0.34 $ & - \\
Cambridge\tablefootmark{d} & 81.5 & $ 13.80 \pm 1.8 $ & - \\
4C\tablefootmark{e}    &  178.0 & $ 5.01 \pm 0.87 $ & - \\
MIYUN\tablefootmark{f} &  232.0 & $ 4.43 \pm 0.05 $ & - \\
WENSS\tablefootmark{g} &  327.0 & $ 3.27 \pm 0.01 $ & $ 1.6 \pm 0.4 $ \\
B2\tablefootmark{h}    &  408.0 & $ 2.30 \pm 0.20 $ & - \\
WSRT\tablefootmark{i}  & 1360.0 & $ 0.76 \pm 0.03 $ & - \\
NVSS\tablefootmark{j}  & 1400.0 & $ 0.750 \pm 0.002 $ & - \\
VLA\tablefootmark{k}   & 1415.0 & - & $ 0.6 \pm 0.1 $ \\
VLA\tablefootmark{l}   & 1490.0 & $ 0.710 \pm 0.003 $ & - \\
BKB77\tablefootmark{m} & 2700.0 & $ 0.37 \pm 0.02 $ & - \\
GB6\tablefootmark{n}   & 4850.0 & $ 0.23 \pm 0.03 $ & - \\
VLA\tablefootmark{o}   & 4885.0 & - & $ 0.20 \pm 0.01 $ \\
\hline 
\end{tabular}
\tablefoot{
\tablefoottext{a}{\cite{CATS, RefWorks:177}}
\tablefoottext{b}{This work}
\tablefoottext{c}{\cite{RefWorks:128}, CLEAN bias correction applied (+0.7 Jy)}
\tablefoottext{d}{\cite{CATS, RefWorks:134}}
\tablefoottext{e}{\cite{RefWorks:242}, flux set to \cite{RefWorks:135} x 1.067 \citep[corrected according to][]{RefWorks:136}}
\tablefoottext{f}{\cite{CATS, RefWorks:137}}
\tablefoottext{g}{\cite{RefWorks:138}}
\tablefoottext{h}{\cite{RefWorks:243}, flux set to \cite{RefWorks:135} x 1.091 \citep[corrected according to][]{RefWorks:136}}
\tablefoottext{i}{This work}
\tablefoottext{j}{\cite{RefWorks:139}}
\tablefoottext{k}{VLA B array, Table \ref{c3:table:1}}
\tablefoottext{l}{VLA C array, \cite{CATS}}
\tablefoottext{m}{\cite{CATS, RefWorks:140}}
\tablefoottext{n}{\cite{CATS, GBT}}
\tablefoottext{o}{VLA C array, Table \ref{c3:table:1}}
}
\end{table}

The best power-law fit to the integrated flux density data points has a spectral index of $ \alpha \, = \, - 1.01 \pm 0.02 $. Although this is not an extremely steep spectral index, it is steeper than what is usually seen in the lobes of radio galaxies ($ \alpha \, \sim \, -0.7 \pm 0.1 $). The integrated spectrum does not show a spectral break.

This indicates that we average over regions of different ages having different intrinsic brightness. The sum of those spectra, each having spectral breaks at different frequencies, produces the remarkably straight integrated spectrum (see the discussion below for details).

\begin{figure*}[htpb]
\centering
\begin{minipage}[c]{0.5\linewidth}
\centering \includegraphics[width=\textwidth]{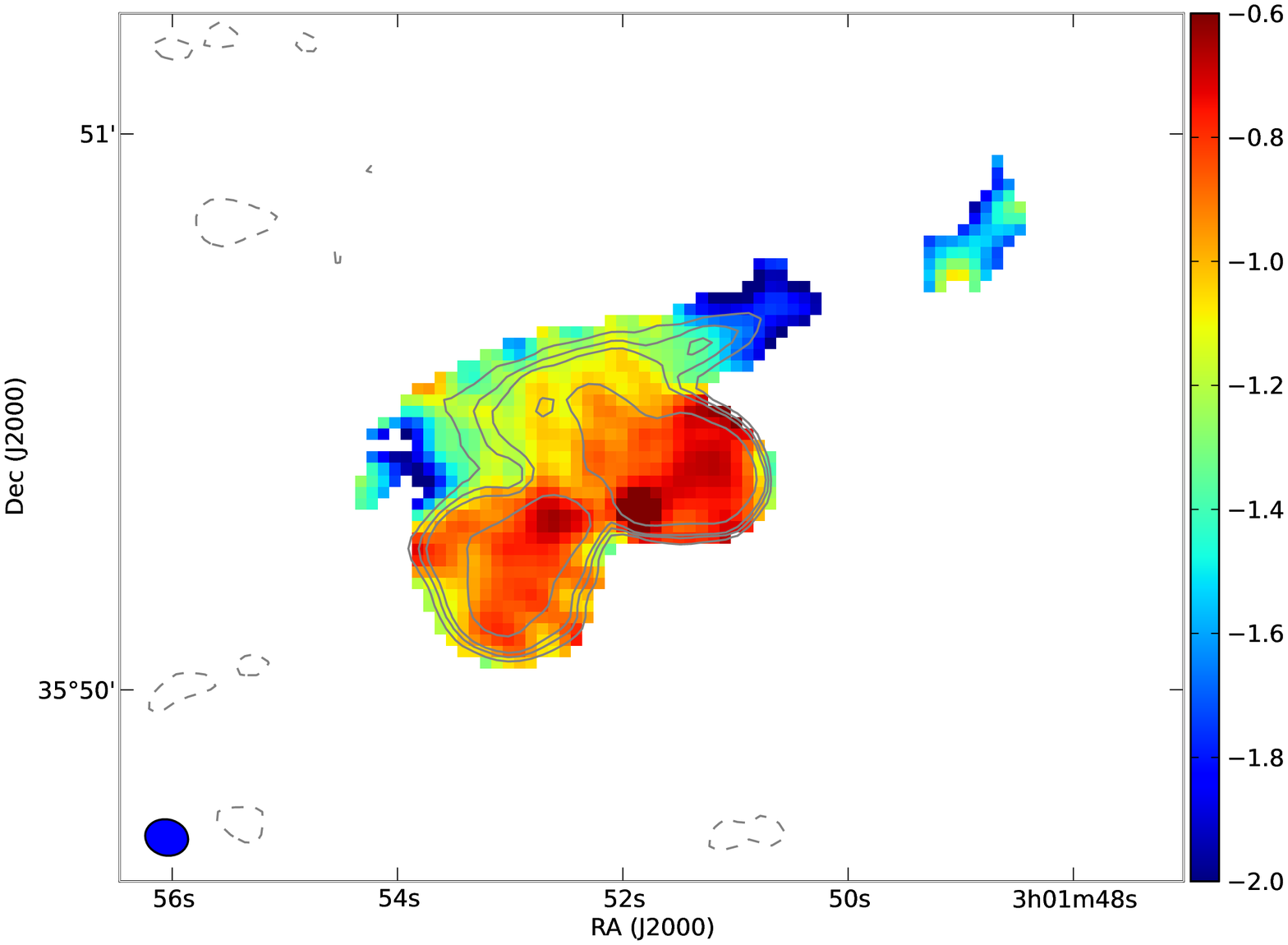}
\end{minipage}%
\begin{minipage}[c]{0.5\linewidth}
\centering \includegraphics[width=\textwidth]{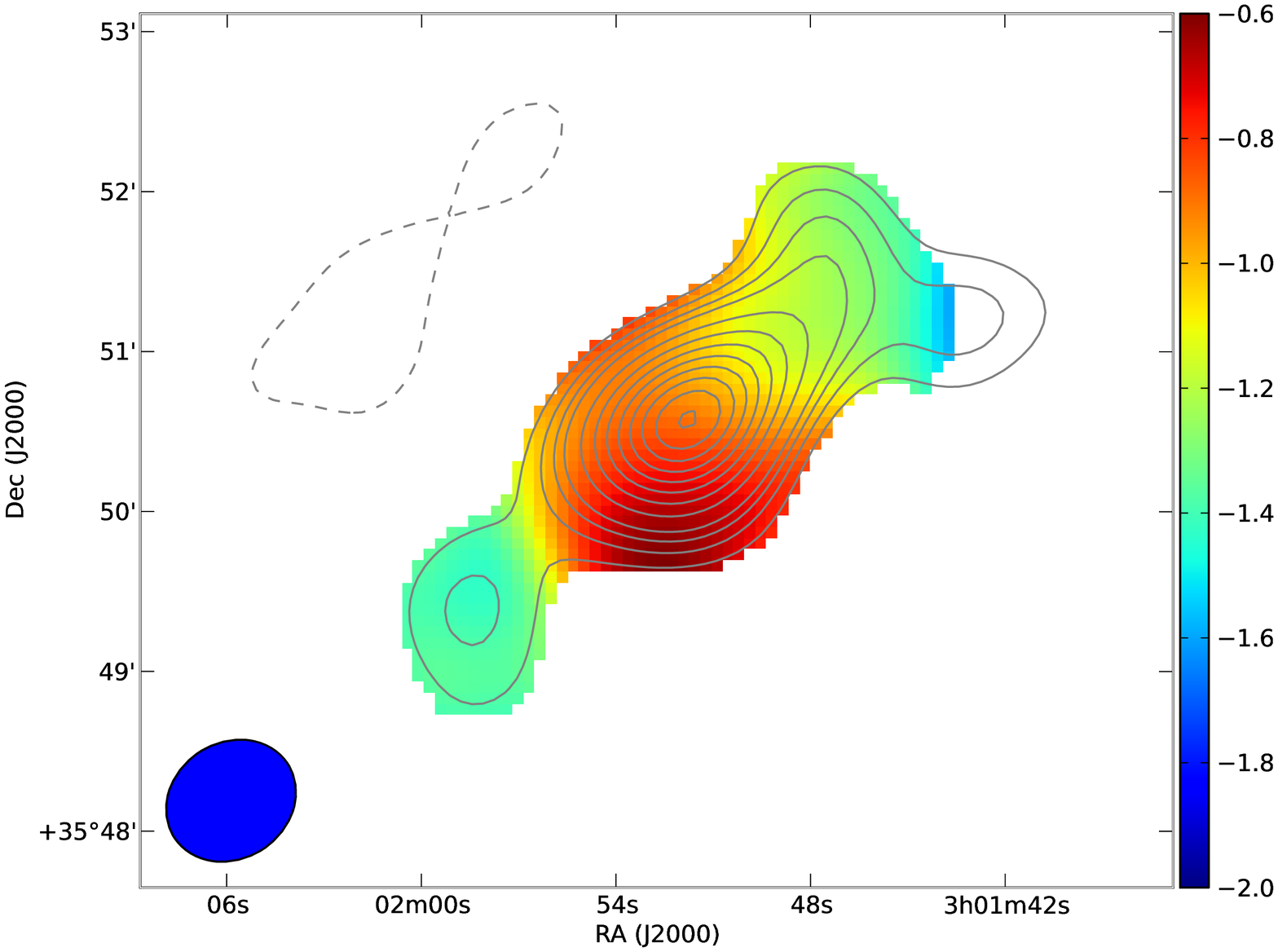}
\end{minipage}
\caption{\textbf{Left:} $ \alpha_{1415}^{4885} $ spectral index map of the inner lobes of 4C~35.06 using the archival VLA images listed in Table \ref{c3:table:1}. Overlaid are VLA 4885 MHz contours $ (-5, 18, 27, 36, 81) \sigma $. \textbf{Right:} $ \alpha_{61}^{1360} $ spectral index map of the extended emission of 4C~35.06 using the LOFAR and WSRT images from this work. Overlaid are 33 contour levels from the LOFAR map spanning the range $ (-3, 183) \sigma $. The beam sizes are given in the lower left corners.}
\label{c3:spix_maps}
\end{figure*}

High-redshift radio sources that will be detected in future radio surveys will be (in most cases) unresolved. Therefore, it is important to benchmark the information obtained from the integrated flux density spectral studies against the one for the spatially resolved studies.

To expand on the different phases of radio-loud activity in this object, we used our LOFAR and WSRT observations, as well as archival VLA images to distinguish between the  spectral index  of the inner regions of 4C~35.06, and the outer regions of extended radio emission. The resulting spectral index images  are shown in Figure \ref{c3:spix_maps}.

The maximum UV range of the LOFAR data set is $ 4 k\lambda $, while that of the WSRT is $ 7 \, k\lambda \times 13 \, k\lambda $. The UV coverage within the range specified for the WSRT is uniform, while the LOFAR data set covers all of the spacings even though it has gaps due to the observation length. We have smoothed the WSRT image to highlight the shorter spacing's. The VLA archival images were obtained from data sets with matching UV ranges and uniform UV coverage.

We see that the inner region where the core and the two brighter lobes are located shows a typical spectral index in the range of $ \alpha \, \sim \, -0.6 $ to $ \alpha \, \sim \, -0.8 $, indicative, as shown by the morphology of the inner double-lobed source, of an active region. However, there is a spectral steepening ($ \alpha \, \sim \, -1.2 $) in the extended emission just to the northwest of the inner lobes (Figure \ref{c3:spix_maps}, left panel). This is the region of the source where we detect the peak flux density in our LOFAR map (see Figure \ref{c3:panel}, right panel), in agreement with LOFAR being more sensitive to regions having steeper spectrum. Thus, this region could indeed represent a region of older radio plasma.

We used lower resolution images to derive a spectral index map for the larger scale regions of the source. The steepening seen at higher resolution in the structure running parallel to the inner brighter lobes, is also visible and continues out to larger scales. This can be seen in the right panel of Figure \ref{c3:spix_maps}. Going to the northwest, the spectral index remains at a value of around $ -1.2 $  up to $ \sim 50 $ kpc from the central region, while on the southeastern side it steepens already to about $ -1.4 $ quite close to the central region. This trend suggests that the outer, extended, regions of the source are composed of aged plasma.

\begin{figure*}[htpb]
\centering
\centering \includegraphics[width=\textwidth]{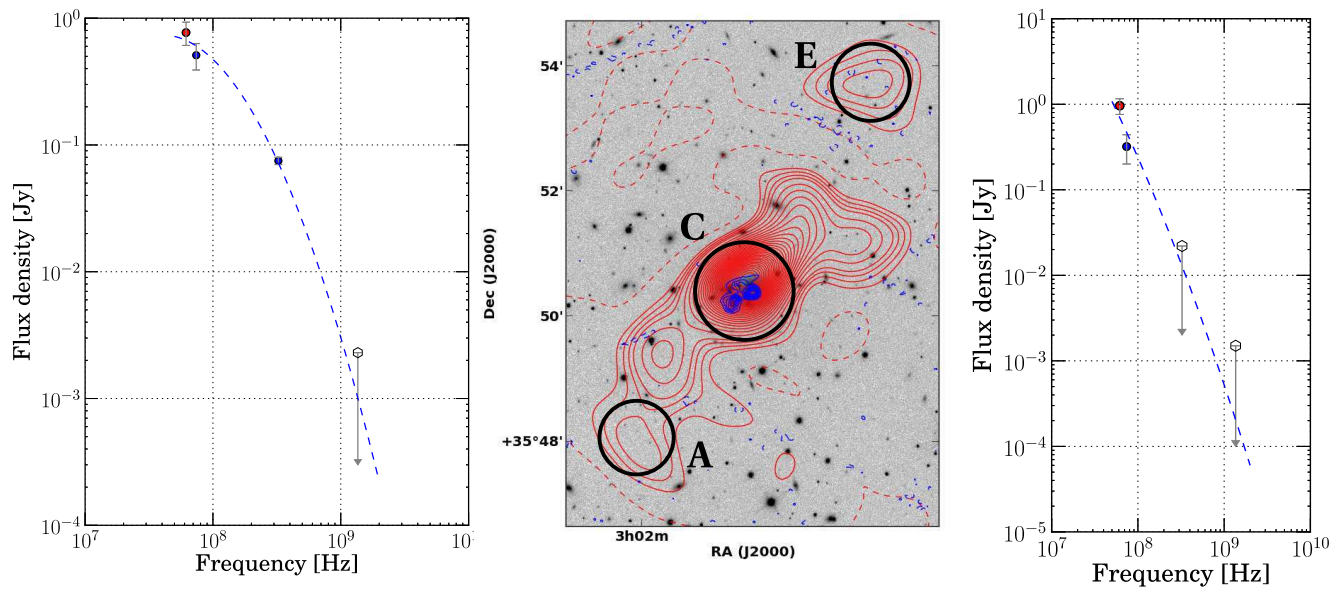}
\caption{\textbf{Left:} The dashed blue line represents a second-order polynomial fit for region A. \textbf{Middle:} The 4C 35.06 radio source as seen by LOFAR, represented by red contours over an inverted grayscale SDSS image of the central region of the Abell 407 cluster. Overlaid in blue are the VLA 4885 MHz image contours. Black circles show the measurement areas in regions A, C, and E. \textbf{Right:} Second-order polynomial fit (dashed blue line) for region E. The data points which are upper limits are represented with empty markers and downward pointing arrows. They are not taken into account in the fitting.}
\label{c3:A_E_spix}
\end{figure*}

We have fitted a synchrotron aging model including a period of activity and a period during which the AGN was switched off \citep[KGJP;][]{RefWorks:188, RefWorks:54} to the integrated flux density measurements at four different frequencies (images listed in Table \ref{c3:table:1}) for the central region of the source (labeled C in Figure \ref{c3:A_E_spix}). We have chosen the KGJP model \citep[Komissarov-Gubanov prescription with a][loss term]{RefWorks:125} since region C contains the radio emission of the currently active AGN as well as an older emission as is evident from the spectral index mapping. We have used an injection spectral index of $ \alpha_{0} \, = \, -0.8 $. The equi-partition value for the magnetic field was calculated to be $ 5.34 \, \mu$G \cite[according to][]{RefWorks:14}. Our best fit model gives a source age of $ t_{\mathrm{s}} \, = \, t_{\mathrm{on}} \, + \, t_{\mathrm{off}} \, = \, 1.5 + 10 $ Myr. Here, $ t_{\mathrm{s}} $ signifies the total source age, comprised of the time the source was active ($ t_{\mathrm{on}} $) and the time elapsed since the AGN shutdown ($ t_{\mathrm{off}} $). The model supports the observation that the central region is the youngest region of the source. We show the fits in Figure \ref{c3:spix}.

We have also determined the spectral index of the outermost source region labeled E, as well as that of region labeled A by measuring the flux density of these regions as seen in the LOFAR image. We have adopted the measurement procedure outlined by \cite{RefWorks:141}. The measurements were done by integrating the flux density over measurement areas in the regions A and E where the surface brightness was above the 3 $ \sigma $ level in the LOFAR, VLSS, WENSS and NVSS images, as well as in our continuum WSRT image and thus calculating the spectral index. The derived values are given in Table \ref{c3:table:3}.

We plot the data from Table \ref{c3:table:3} in Figure \ref{c3:A_E_spix} and fit a power-law to the data points for region A. The spectrum of region E is poorly constrained, but we note that it is ultra steep ($ \alpha < -2 $). Region A has a flatter spectrum ($ \alpha \, = \, -1.36 \pm 0.09 $) in line with what we expect from the spectral trend of the inner regions. More sensitive observations at higher frequencies (with sufficient resolution) are needed to determine the exact spectral index of the outermost regions. In Figure \ref{c3:A_E_spix} we fit a second-order polynomial to the measurements to illustrate the aged spectra.

\begin{table}[!]
\caption{\small Measured flux density values for the regions labeled A and E in the LOFAR image.}
\label{c3:table:3}
\centering
\begin{tabular}{p{2cm} p{2.5cm} p{2.2cm}}\\     
\hline\hline\\
$ \nu $ [MHz] & $ S_{\mathrm{\nu}} $ [Jy] & \\
\hline\\
 & Region A & $ \alpha $ \\
\hline\\
  61.6 & $ 0.77 \pm 0.16 $ & $ -1.36 \pm 0.09  $ \\
  74.0 & $ 0.51 \pm 0.12 $ & \\
 325.0 & $ 0.075 \pm 0.005 $ & \\
1360.5 & $ < 0.0023 $ & \\
\hline\\
 & Region E & $ \alpha_{62}^{74} $ \\
\hline\\
  61.6 & $ 0.96 \pm 0.20 $ & $ < \, -2 $\tablefootmark{a}\\
  74.0 & $ 0.32 \pm 0.12 $ & \\
 325.0 & $ < 0.022 $ & \\
1360.5 & $ < 0.0015 $ & \\
\hline
\end{tabular}
\tablefoot{
\tablefoottext{a}{Region E is detected in the LOFAR and VLSS images, and the other measurements are only upper limits. Hence, we only show an illustrative polynomial fit in Figure \ref{c3:A_E_spix}, but do not attempt to derive a spectral index.}
}
\end{table}

\section{Discussion}
\label{c3:disc}

Figure \ref{c3:VLA_in} shows the radio emission around the galaxies comprising the Abell~407 core (cataloged as UGC~2489). We can see that there are two lobes detected by the VLA, and diffuse emission that extends to the northeast. Going farther out, this emission extends into a helical extension as seen in the grayscale WSRT image shown in the left panel of Figure \ref{c3:panel}.

The extended emission to the northwest continues out to an even larger distance than previously known, i.e. $ \sim 210 $ kpc from the central region of 4C~35.06. It is labeled as region E in the LOFAR image (see Figure \ref{c3:panel}).

The overall source morphology is suggestive of (at least) two distinct episodes of radio activity in 4C~35.06. The extended emission is a remnant of the previous active phase, whilst the twin lobes detected by the VLA show the current activity cycle.

Using the morphological and spectral results described above, we propose that 4C~35.06 can be interpreted as a source in which we are seeing (at least) two subsequent episodes of AGN activity.

To explain the complex radio morphology as well as the spectral indices we have derived, we propose that the AGN is intermittently active as it moves in the dense environment in the cluster core. In this scenario, the AGN turned on sometime in the past, and has produced the helical pattern of emission visible in the left panel of Figure \ref{c3:panel}, possibly a sign of jet precession / merger during that episode of activity. Using LOFAR, we can trace the relic plasma from that episode of activity out to greater distances from the core than ever before. The spectral index maps of the source indicate that this phase of activity has resulted in an FRI-type radio galaxy (spectrum steepening from the core toward the edges). Then, the AGN activity has ceased (perhaps gradually), and after a period of time during which the host galaxy moved some distance to the position we observe it at now, it has started again, producing the less steep spectrum inner lobes. These lobes have clear double morphology, they are clearly outlined and both morphologically and spectrally distinct from the surrounding steeper spectrum region. At the present epoch, we observe an aged FRI-like large scale structure with an embedded restarted radio source, similar to the objects analyzed for example by \cite{RefWorks:34}, \cite{RefWorks:250} and \cite{RefWorks:127}.

An alternative possibility would be that the different radio components (on small and large scales) belong to activity connected to different cores of UGC~2489 that were active at different times. For example, a core that could have been active in the past is the core labeled E in \cite{RefWorks:111}. It is located (in projection) to the northeast of the currently active AGN host core, and is co-spatial with the ridge of radio emission which continues out to larger scales. \cite{biju} identify it as the origin of (at least) the large scale extended emission (they do not comment on the double lobed source morphology indicating restarted AGN activity to the southwest). The steeper spectral index of the ridge of radio emission, indicating old plasma (compared to the currently active double lobes just to the southwest of it) raise the possibility that this core hosted an active AGN sometime in the past and is responsible for the component of radio emission that is seen, in projection, associated with 4C~35.06. In our opinion, this is less likely; the optical core in question is less massive (judging by its optical luminosity compared to other nearby cores) than the currently active one and represents a less likely host of a SMBH.

We can estimate the AGN activity timescale by using galaxy kinematics. The cores of UGC~2489 are embedded in a common stellar envelope (probably a remnant of previous mergers). We assume that the core hosting the AGN is moving within the envelope and that the AGN is intermittently active. The radial velocity difference of the AGN host galaxy and the stellar envelope \cite[as reported by][]{RefWorks:111} is 320 \kms. Assuming that the radial velocity is equal to the velocity in the plane of the sky, the projection of the section of the orbit traversed can be estimated by measuring the current angular separation between the AGN host galaxy and the "ridge" of steeper spectrum emission to the northeast. This separation is 13\arcsec\ or 11.3 kpc at the distance of UGC~2489. Under the assumption that the shutdown and restart were near instantaneous, we get a lower limit for the elapsed time between the two episodes of AGN activity of around $ 35 $ Myr. It matches the duration of the AGN dormant phase of some of the fading radio sources reported in the sample of \cite{RefWorks:96}, and also is in agreement with the derived duration of the inactive phases for the sample of \cite{RefWorks:34}. It is larger than the off times found (which are on the order of several Myr) for the two double-double radio galaxies studied by \cite{RefWorks:178}.

We have been unable to unambiguously determine a spectral break in the integrated source spectrum. Furthermore, the images we have lack the proper frequency coverage at high spatial resolution to allow us to determine a spectral break and arrive at a radiative age estimate for the outermost emission regions.

We can estimate at what frequency would the spectral break be located for the aged plasma originating from the past activity episode (before the AGN shutdown). Following \cite{RefWorks:34}, the break frequency $ \nu_{\mathrm{b}} $[GHz] is related to the magnetic field strength in the emission region $ B $[$ \mu $G] and the time elapsed $ t_{\mathrm{s}} $[Myr] since the start of the activity as

\[ \nu _{\mathrm{b}} = \frac{1}{1 + z} \cdot \left( 1590 \cdot \frac{B^{1/2}}{(B^{2} + B_{\mathrm{IC}}^{2}) \cdot t _{\mathrm{s}}} \right) ^{2} \]

\noindent where $ B_{\mathrm{IC}} = 3.25(1 + z)^{2} $ is the inverse Compton equivalent magnetic field. We assume (for simplicity) that the active and shut down timescales of the AGN are the same, $ t _{\mathrm{s}} = 70 $ Myr in our case. Taking Region A as an example, we can estimate the value of its equi-partition magnetic field according to \cite{RefWorks:24}. For an electron to proton energy ratio of unity and a Lorentz factor of the lowest energy electrons of $ \gamma = 700 $, the magnetic field estimate is $ 2.8 \mu $G, with a corresponding break frequency of $ \nu_{\mathrm{b}} \, = \, 3.3 $ GHz. However, the break frequency value depends critically on the value of the magnetic field which can vary substantially. For example, for the lowest values of the Lorentz factor, the magnetic field can be as high as $ 10 \mu $G, with a corresponding spectral break frequency of $ 380 $ MHz. Such a value for the break frequency is plausible if we look at the spectral shape of Region A shown in Figure \ref{c3:A_E_spix}, left panel.

Region E has a steeper spectrum than region A. This observation, (as well as the source morphology) is supported by the preliminary measurement of its flux density at 150 MHz using LOFAR (Marisa Brienza, private communication). It may mean that region E is older; it may actually be the oldest source region that we observe.

\cite{RefWorks:111} estimate that the UGC~2489 galaxies will merge owing to dynamical friction (as shown by the stellar envelope surrounding the galaxies, created by stripping) in at most 2 Gyr, leaving us the possibility for the AGN to experience a series of outbursts before the final merger.

An interesting similar example is presented by the radio source 3C~338, hosted by the cD galaxy (NGC~6166) in the core of the galaxy cluster Abell~2199. This is a steep spectrum radio source ($ \alpha < -1.5 $). The radio morphology \citep{RefWorks:127} is suggestive of a restarted source, with at least two phases of AGN activity, where the oldest radio plasma appears to be offset from an active core. In these respects, this radio source is very similar to 4C~35.06.
However, 3C~338 is different from 4C~35.06 in that the older plasma in its central regions has no optical core counterpart, which gives a stronger support to the argument that in the case of 3C~338 the currently active core was the source for the older plasma. \cite{RefWorks:159} give a thorough overview of radio sources hosted by cluster core galaxies and suggest how the motion of the AGN host galaxy can give rise to a radio morphology similar to what we observe. We note that merging of galaxy cores on timescales of $ 10^{7} $ yrs. can trigger an activity episode, which is in agreement with the derived dormant time for 4C~35.06.

In these types of restarted radio sources which are located in galaxy clusters, there is a possibility that the ICM medium is dense enough near the cluster core, increasing confinement pressure. It can have an impact on any older outburst of AGN plasma, keeping it confined for longer periods of time, and prolonging the time in which it can be detected by low-frequency radio observations such as we are presenting in this work.

\subsection{Presence of cold gas}
\label{c3:coldgas}

The detection of neutral hydrogen adds a new interesting component to the study of this object.

\HI\ in absorption has been detected before in clusters \citep[for example Abell~2597 (PKS~2322-123)][]{RefWorks:166, RefWorks:167}, Abell~1795 \citep[4C~26.42,][]{RefWorks:168} in addition to other famous objects in centers of clusters, like Hydra~A \citep{RefWorks:173} and Cygnus~A \citep{RefWorks:169}. Although a systematic inventory of the occurrence of \HI\ in clusters is not yet available, the statistics seems to be biased toward observing and detecting \HI\ in the center of cooling core clusters. Abell~407 likely does not belong to this group,  having a core which is not relaxed, but nevertheless we detect \HI.

The fact that 4C~35.06 is a restarted radio source may suggest a connection between the presence of gas and the restarting of the radio-loud phase: indeed such a relation is often seen in non-cluster radio sources. There, restarted radio sources show an  incidence of \HI\ detections higher than in other extended sources \citep{RefWorks:170, RefWorks:89, RefWorks:171, RefWorks:132}.

However, the \HI\ profile detected in 4C~35.06 appears to be broader than in typical radio galaxies \citep{RefWorks:249}. However, it resembles some of the profiles found in other clusters (in particular Abell~2597 and Abell~1795 mentioned above) that show relatively broad wings in the \HI\ absorption profile, perhaps a signature of the more disturbed kinematics of the gas due to the cluster environment, as seen also in molecular gas by \cite{RefWorks:165} derived from ALMA data of Abell~1835.

In low-luminosity radio sources with low-excitation spectra  like 4C~35.06, the possibility of the activity being supported by gas cooling from the hot galactic halo via e.g. Bondi accretion has been suggested by several studies \citep{RefWorks:70, RefWorks:75, RefWorks:62, RefWorks:95}. Taking the core luminosity measured by \cite{RefWorks:110} and following the reasoning of \cite{RefWorks:95}, connecting the radio core luminosity to the radio jet power and relating that to the Bondi accretion power ($ P_{\mathrm{B}}=0.1\dot{M}c^{2} $), we estimate that the accretion rate of 4C~35.06 is on the order of 0.2 $ M _{\odot} yr ^{-1} $.

If part of the detected \HI\ is associated with infalling gas (something possible given the width of the profile that also spans to redshifted velocities) it could provide the required fueling material. The gas could be chaotically losing angular momentum in the ISM of the AGN host and accreting, as suggested by \cite{RefWorks:143}.

\section{Conclusions}
\label{c3:fin}

Studying the field centered on the nearby Compact Steep Spectrum radio source B2~0258+35A (observed during the commissioning of LOFAR), we have made a 61 MHz map of 4C~35.06 (B2~0258+35B), a radio source hosted by the core of the galaxy cluster Abell~407. This source has a complex morphology on several spatial scales. We used our LOFAR image in combination with archival VLA imaging as well as VLSS, WENSS and other survey data to argue that the most extended emission can be interpreted as a remnant of a past episode of AGN activity. The integrated spectral index is moderately steep and shows no signs of spectral curvature. We confirm the existence of an extended emission region and derive spectral index maps of the inner parts of the source closer to the AGN.

We discuss two possible scenarios regarding the nature of the source and conclude that the best explanation is that we observe two distinct episodes of AGN activity in 4C~35.06. Fitting a radiative ageing model to the spectrum of the central source region (encompassing the currently active AGN as well as aged plasma) we conclude that it is at most $ 10 $ Myr old, while age estimates (based on galaxy kinematics and radio spectral shape) for a source region farther away from the AGN show it to be older by at least a factor of a few. Although the outer region age estimates are less precise, they support the suggestion of multiple phases of activity.

We also present a discovery of \HI\ in absorption at the location of the (currently active) AGN host galaxy. The \HI\ detection has a broad profile, suggestive of a complex gas dynamics and a possible outflow.

A better handle of the timescales involved can be obtained by better spectral coverage at sufficient resolution to disentangle the various source structures. Depending on the relative contributions of the smaller scales, our study shows that it is possible to use the integrated source spectrum as a proxy to learn more about the spectral behavior on smaller scales and the source activity history. This is important in studies of more distant sources which in general would be unresolved in future surveys. It is usually assumed that relic sources at high redshift should have very steep spectra, since we observe the exponential cutoff in their spectra which in our frame of reference is shifted to lower frequencies. However, for some of these sources, restarted activity may modify their (integrated) spectral shapes such that they would be less steep. An example of such a case is presented by the integrated flux density spectrum of 4C~35.06, the source we have studied in this work. Our observations suggest that we have to be careful and possibly modify our radio selection criteria of high-redshift sources.

\begin{acknowledgements}

LOFAR, the Low Frequency Array designed and constructed by ASTRON, has facilities in several countries that are owned by various parties (each with their own funding sources), and that are collectively operated by the International LOFAR Telescope (ILT) foundation under a joint scientific policy.\\
Data taken with the Westerbork Synthesis Radio Telescope were part of this work. It is operated by ASTRON (Netherlands Institute for Radio Astronomy) with support from the Netherlands Foundation for Scientific Research (NWO).\\
This research has made use of the NASA/IPAC Extragalactic Database (NED), which is operated by the Jet Propulsion Laboratory, California Institute of Technology, under contract with the National Aeronautics and Space Administration.\\
The VLA images used in this work were produced as part of the NRAO VLA Archive Survey, (c) AUI/NRAO.\\
This research has made use of APLpy, an open-source plotting package for Python hosted at http://aplpy.github.com.\\
RM gratefully acknowledges support from the European Research Council under the European Union's Seventh Framework Programme (FP/2007-2013) / ERC Advanced Grant RADIOLIFE-320745.\\
RJvW acknowledges the support by NASA through the Einstein Postdoctoral grant number PF2-130104 awarded by the Chandra X-ray Center, which is operated by the Smithsonian Astrophysical Observatory for NASA under contract NAS8-03060.

\end{acknowledgements}

{}

\end{document}